

\input harvmac.tex

%
\def\ubrackfill#1{$\mathsurround=0pt
	\kern2.5pt\vrule depth#1\leaders\hrule\hfill\vrule depth#1\kern2.5pt$}
\def\contract#1{\mathop{\vbox{\ialign{##\crcr\noalign{\kern3pt}
	\ubrackfill{3pt}\crcr\noalign{\kern3pt\nointerlineskip}
	$\hfil\displaystyle{#1}\hfil$\crcr}}}\limits
}

\def\ubrack#1{$\mathsurround=0pt
	\vrule depth#1\leaders\hrule\hfill\vrule depth#1$}
\def\dbrack#1{$\mathsurround=0pt
	\vrule height#1\leaders\hrule\hfill\vrule height#1$}
\def\ucontract#1#2{\mathop{\vbox{\ialign{##\crcr\noalign{\kern 4pt}
	\ubrack{#2}\crcr\noalign{\kern 4pt\nointerlineskip}
	$\hskip #1\relax$\crcr}}}\limits
}
\def\dcontract#1#2{\mathop{\vbox{\ialign{##\crcr
	$\hskip #1\relax$\crcr\noalign{\kern0pt}
	\dbrack{#2}\crcr\noalign{\kern0pt\nointerlineskip}
	}}}\limits
}

\def\ucont#1#2#3{^{\kern-#3\ucontract{#1}{#2}\kern #3\kern-#1}}
\def\dcont#1#2#3{_{\kern-#3\dcontract{#1}{#2}\kern #3\kern-#1}}



\font\tenmsy=msbm10
\font\sevenmsy=msbm10 at 7pt
\font\fivemsy=msbm10 at 5pt
\newfam\msyfam 
\textfont\msyfam=\tenmsy
\scriptfont\msyfam=\sevenmsy
\scriptscriptfont\msyfam=\fivemsy
\def\blackB{\fam\msyfam\tenmsy}
\def\ZZ{{\blackB Z}}

\def\II{{\blackB I}}

\let\d\partial

\let\s\sigma
\let\L\Langel
\let\R\rangle

\def\frac#1#2{{\textstyle{#1\over #2}}}

\def\eqalignD#1{
\vcenter{\openup1\jot\halign{
\hfil$\displaystyle{##}$~&
$\displaystyle{##}$\hfil~&
$\displaystyle{##}$\hfil\cr
#1}}
}

\def\text#1{\quad\hbox{#1}\quad}

\def\P{{\cal {P}}}

\def\y{{\infty}}

\def\rw{\rightarrow}

\def\L{\langle}
\def\R{\rangle}

\def\Y{{\cal Y}}

\newcount\eqnum
\eqnum=0
\def\eq{\eqno(\secsym\the\meqno)\global\advance\meqno by1}
\def\eqlabel#1{{\xdef#1{\secsym\the\meqno}}\eq }

\newwrite\refs
\def\startreferences{
 \immediate\openout\refs=references
 \immediate\write\refs{\baselineskip=14pt \parindent=16pt \parskip=2pt}
}
\startreferences

\refno=0
\def\aref#1{\global\advance\refno by1
 \immediate\write\refs{\noexpand\item{\the\refno.}#1\hfil\par}}
\def\ref#1{\aref{#1}\the\refno}
\def\refname#1{\xdef#1{\the\refno}}

\newcount\exno
\exno=0
\def\Ex{\global\advance\exno by1{\noindent\sl Example \the\exno:

\nobreak\par\nobreak}}

\parskip=6pt

\overfullrule=0mm

\def\frac#1#2{{#1 \over #2}}

\let\d=\partial

\def\suh{{\widehat {su}}}
\def\uh{{\widehat u}}
\def\rw{{\rightarrow}}

\newwrite\refs
\def\startreferences{
 \immediate\openout\refs=references
 \immediate\write\refs{\baselineskip=14pt \parindent=16pt \parskip=2pt}
}
\startreferences

\refno=0
\def\aref#1{\global\advance\refno by1
 \immediate\write\refs{\noexpand\item{\the\refno.}#1\hfil\par}}
\def\ref#1{\aref{#1}\the\refno}
\def\refname#1{\xdef#1{\the\refno}}


\Title{\vbox{\baselineskip12pt
\hbox{
}}}
{\vbox {\centerline{Single-channel correlators and  residue calculus }}}

\smallskip
\centerline{ P. Jacob and P. Mathieu
}

\smallskip\centerline{ \it D\'epartement de
physique,} \smallskip\centerline{Universit\'e Laval,}
\smallskip\centerline{ Qu\'ebec, Canada G1K 7P4}
\smallskip\centerline{(pjacob@phy.ulaval.ca, pmathieu@phy.ulaval.ca)}
\vskip .2in
\bigskip
\bigskip
\centerline{\bf Abstract}
\bigskip
\noindent
Some simple (namely, single-channel) correlation functions involving an arbitrary number of
fields are computed by means of a direct application of the residue calculus,  through  partial
fraction expansions. Examples are presented in minimal models and  parafermionic conformal theories. A
generic factorized expression is deduced  for the corresponding single-channel structure constants.


\Date{05/01\ }


\newsec{Introduction}

\subsec{Correlators, OPEs and Ward identities }

The plain method for evaluating $N$-point correlation functions, given the OPEs (operator product
expansions) of the fields whose correlations are to be computed, amounts to substitute the OPE of two
fields in order to reduce the correlator to
$(N-1)$-point functions and iterate this procedure until the result becomes expressed in terms
of a three-point function. The later being known exactly up to structure constants, the
correlator is then expressed in terms of these constants. However, the
reduction in the number of points has been traded for a new complication: we then
have to sum up the infinite series (i.e., conformal blocks) associated to each OPE. In principle, these
can be summed exactly only in simple cases. A sample computation is presented in Appendix A. 

 For the mere formulation of the model, the correlators that are particularly important are
those involving the  symmetry generators, namely the generators of the extended
conformal algebra. Their relevance lie  in that the internal coherence of the
extended algebra  boils down to a precise statement concerning the correlators of the
symmetry generators: all their four-point functions must be {\it associative}. 
The associativity requirement is the condition that a correlation
function can be calculated in many different ways, in particular, by evaluating the
OPEs in different orders, without affecting the result. (This is the way the structure
constants are calculated.) But testing associativity calls for the {\it exact} form of these
correlation functions. 

However, in this particular instance, the problem appears to be tractable. 
Indeed, the evaluation of correlation functions
involving extended conformal-algebra generators is usually rather simple in that only
the {\it singular terms} have to be considered. Take for instance  a correlation function involving the
energy-momentum tensor $T(z)$ and some primary fields $\prod_{j=1}^N \phi_j(z_j)$. This is
certainly relevant  to the question of studying the associativity of a given conformal
algebra because the algebra generators other that $T$ have to be Virasoro primary fields. It is
a very basic fact that in eliminating
$T(z)$ through OPE, one simply needs to
take into account the singular terms in the product of
$T(z)$ with all the other fields in the correlator:
$$\L T(z) \prod_{j=1}^N \phi_j(z_j)\R = \sum_{i=1}^N\left\{ {h_i\over
(z-z_i)^2}+{\d_{z_i}\over z-z_i}\right\}\L  \prod_{j=1 }^N
\phi_j(z_j)\R\eqlabel\virwardd$$
 In other words, in this special case we do not have to
keep track of an infinite series. This is a consequence of the {\it conformal Ward
identities} - cf. [\ref{A.A. Belavin, A.M. Polyakov and A.B. Zamolodchikov,
Nucl. Phys. {\bf B241} (1984) {333}.}\refname\BPZ, \ref{A.L. Zamolodchikov, Theo. Math. Phys. {\bf 63}
(1985) 1205.}\refname\Zam, \ref{P. Di Francesco, P. Mathieu and D. S\'en\'echal, {\it Conformal Field
theory}, Springer Verlag, 1997.}\refname\CFT]. 

\subsec{Correlators involving $T$ as meromorphic functions }

But if we think about this result from the point of view of OPEs, it looks rather surprising
that  by considering only the singular terms of the OPE of
$T$ with the other fields of the correlator (and not the complete infinite OPE series) we
can compute the correlation function exactly.  
 This has a natural complex-analysis explanation: the result simply corresponds to the 
partial fraction expansion of the meromorphic function representing the correlation,
viewed as a function of one of its field.   For instance, the correlator in (\virwardd),
considered as a function of
$z$, is a meromorphic function with double poles at the various $z_j$ (since the
$\phi_j(z_j)$'s are supposed to be primary) with coefficients fixed by the OPE $T(z)\phi_j(z_j)$.
In that case, the partial fraction expansion is complete in that there is no
additional analytic piece. Indeed, this meromorphic function vanishes at infinity since
$T(z)\sim z^{-4}$ as
$z\rw
\y$. Such a function is simply given by the sum of
the principal parts at the various poles.

Let us make the above statements more explicit. Recall that a meromorphic function with vanishing
analytic part can  be written in  partial
fraction as
$$F(z) = \sum_{j=1}^N \sum_{r=1}^{n_j}{a_{r}^{(j)}\over (z-z_j)^{r}}\eq$$
where
 $n_j$ is the order
of the pole at $z_j$ and the coefficients
$a_r^{(j)}$ are given by
$$a_{n_j-k}^{(j)}= \lim_{z\rw z_j}{1\over k!}{\d^k\over \d z^k}(z-z_j)^{n_j}
F(z)\eqlabel\ctea$$ 
When $F(z)$ is a correlation function, the various coefficients
$a_r^{(j)}$ are expressed in terms of lower-order correlation functions. 
On the other hand, if the meromrophic function $F(z)$ does not vanish at infinity, an analytic function
needs to be added to this sum of principal parts.  Suppose that $F(z)$  behaves rather like
$F(z)\sim z^p$ as $z\rw \y$, with $p$ integer. This signals the presence of a pole of order $p$ at
infinity so that the principal part at infinity (in which we include a constant term) has to be taken
into account. The expression of $F(z)$ becomes then
$$F(z) = \sum_{j=1}^N \sum_{r=1}^{n_j}{a_{r}^{(j)}\over (z-z_j)^{r}}+ \sum_{k=1}^{p}a_{k}^{(\y)}\, z^k
\eq$$
The last sum is the analytic part of $F(z)$.

It is then completely obvious that in the case (\virwardd), the principal part at
$z_j$ is nothing but the sum of the singular terms in the OPE $T(z)\phi_j(z_j)$. Indeed, viewing the
correlator as a function of $z$, we have:
$$\L T(z) \prod_{j=1}^N \phi_j(z_j)\R= \sum_{i=1}^N \left({a^{(i)}_2\over
(z-z_i)^2}+{a^{(i)}_1\over (z-z_i)}\right)\eq$$
with
$$\eqalign{
a^{(i)}_2 &= \lim_{z\rw z_i}(z-z_i)^{2} \L T(z) \prod_{j=1}^N \phi_j(z_j)\R \cr
& = \lim_{z\rw z_i}(z-z_i)^{2} \L  \prod_{j=1}^{i-1} \phi_j(z_j)\left({h_i\phi_i(z_i)\over
(z-z_i)^2}+{\d_{z_i}\phi_i(z_i)\over z-z_i}+\cdots \right) \prod_{j=i+1}^{N} \phi_j(z_j)\R
\cr&= h_i\L  \prod_{j=1 }^N
\phi_j(z_j)\R\cr
}\eq$$
and
$$\eqalign{
a^{(i)}_1&=   \lim_{z\rw z_i}{\d\over \d z} (z-z_i)^{2} \L T(z) \prod_{j=1}^N \phi_j(z_j)\R \cr
& = \lim_{z\rw z_i} {\d\over \d z} \left\{ \L h_i \prod_{j=1}^{N}
\phi_j(z_j)\R+ (z-z_i) \d_i \L  \prod_{j=1}^{N}
\phi_j(z_j)\R+\cdots \right\}
\cr &= \d_i\L  \prod_{j=1 }^N
\phi_j(z_j)\R\cr}\eq$$
(the first term of the second line drops because it is independent of $z$); this gives the rhs
of (\virwardd).

The above derivation of (\virwardd) makes clear the rather auxiliary  aspect of the
primary nature of the fields inside the correlator. For instance, one could introduce 
quasi-primary fields, e.g., $T(z)$ itself. Regarding the following correlator as a function
of $\zeta_1$ leads to [\Zam]:
$$\eqalign{
&\L \prod_{i=1}^M T(\zeta_i) \prod_{j=1}^N \phi_j(z_j)\R \cr
&\qquad  = \sum_{i=2}^M {c/2\over (\zeta_1-\zeta_i)^4}\L \prod_{k=2\atop k\not=i}^M
T(\zeta_k)
\prod_{j=1}^N
\phi_j(z_j)\R \cr
 &\qquad   + \sum_{i=2}^M \left({2\over (\zeta_1-\zeta_i)^2}+{1\over
\zeta_1-\zeta_i}{\d\over \d \zeta_i}\right)\L \prod_{k=2}^M T(\zeta_k)
\prod_{j=1}^N
\phi_j(z_j)\R \cr
&\qquad   + \sum_{i=1}^N \left({h_i\over (\zeta_1-z_i)^2}+{1\over
\zeta_1-z_i}{\d\over \d z_i}\right)\L \prod_{k=2}^M T(\zeta_k)
\prod_{j=1}^N
\phi_j(z_j)\R \cr}\eq$$
A manifestation of the associativity property  is that the same expression for this
correlator follows by considering it as a meromorphic function of another variable $\zeta_i$.

 Even `less-primary' fields are treated by the same method: only the
singular terms of the OPE with $T(z)$  contribute.


\subsec{Generalizing the class of correlators computable from partial fraction expansions }

What is central in the previous computations of correlators involving $T$? At first, there is
the fact that the OPE of
$T$ with another field has only one channel, i.e., all those terms that appear in the OPE
$T(z)\phi(w)$ belong to the conformal family of $\phi(w)$. This ensures that the different
powers of $z-z_j$ (for a given $j$) all differ by integers. A second important point is that
$T$ is local with respect to any other field in the theory, which implies that, not only
the different
powers of $z-z_j$ differ by integers, but they are all integers themselves. In
other words, the OPE
$T(z)\phi(w)$ is a genuine Laurent series. 

Therefore, this computation method can be used for calculating the different four-point
functions for the extended-algebra generators when these generators  $\Y_i$ all have integer dimensions.
Note that in this case, the one-channel constraint
is superfluous because all the powers of $z-w$ in the OPE  $\Y_i(z)\Y_j(w)\in \sum_k[\Y_{k}(w)]$ have
automatically integer dimension.  A well-known example of this type is the  WZW model whose extended
symmetry is an affine Lie algebra, with generators
$J^a$ satisfying the OPE $J^a\times J^b= \II+J^c$.

But if this is to be used for the  analysis of more general extended algebras, for instance parafermionic
models with fields having fractional dimension, one has to be able to tackle situations
where there are branching points, that is, when the powers of $z-w$ are fractional in the
OPE 
$\Y_i(z)\Y_j(w)$. The cure is quite simple:  we just have to modify the correlation
function, which is
 viewed as
a function of the first variable $z_1$ -- the position of a given symmetry generator,
say $\Y(z_1)$ --,  by multiplying it by appropriate
powers of
$z_1-z_j$ ($z_j$ being the position of another field inside of the correlator) in order
to transform it into a meromorphic function of
$z_1$ [\ref{A.B.
Zamolodchikov and V.A. Fateev, Sov. Phys. JETP {\bf 43} (1985) 215.}\refname\ZFa].  In other words, 
to analyze the correlation $\L 
\Y(z_1)\cdots\R$, we consider the
intermediate function
$$F(z_1)= \left(\prod_{j\geq 2} (z_{1j})^{d_j} \right) \L\Y(z_1)\cdots\R\eq$$
for those values of $d_i$ appropriate to make $F(z_1)$  meromorphic.
The OPEs fix then the position of
the poles of this function together with their residues.  This determines  $F(z_1)$ and, thereby,
the correlation function under consideration. Notice however that in the present case, the principal
part of $F(z_1)$ at  $z_j$ is not given solely by the sum of the singular terms in
$z_1-z_j$: for the subleading terms in the principal series, 
there are derivatives in the expression for the constants $a^{(j)}_{n_j-k}$ in (\ctea) and these
derivatives do not select exclusively the corresponding singular terms due to the presence of the
prefactors
$\prod_{i\geq 2} (z_{1i})^{d_i}$ on which the derivatives also act.

As already pointed out, the presence or absence of a regular (analytic) part in the expression
of
$F(z_1)$ is fixed by the behavior of the field at position $z_1$,  as $z_1\rw \y$,
e.g.,
$ \Y (z_1)\sim z_1^{-2h_{\Y}} $, together with the prefactor composed of the different
fractional powers of
$z_1-z_j$  that have been introduced.  A modification of the prefactor by an integer power of $z_{1j}$
obviously affects the nature of the regular part of
$F(z_1)$ but not the final expression of the correlator.

By bringing out the conditions underlying the applicability of the residue method to the computation of
correlators involving $T$, we have identified at once two criteria: (1) the single-channel requirement
and (2) the locality condition. 
The simple trick just described for transforming a function with branching
singularities into a meromorphic function provides thus a way of bypassing the apparent second
limitation.

In relation with the single-channel requirement, we have already presented a situation in which it can
be relaxed, namely for correlators of symmetry generators all having integer dimension. Phrased in
more general terms,  the one-channel condition is not mandatory  when the OPEs of the fields inside
the correlators close in a set of fields which all have conformal dimension that differs from each
other by integers.  But this is a rather special instance. In more general circumstances, the
single-channel requirement cannot be avoided. Notice however that this
is not as restricting as it looks at first sight. What is really needed is a not exactly a genuine
single-channel OPE but, rather, a correlation function that selects a single channel in each 
intermediate OPE.

In the above considerations there is in addition an implicit third limitation, which is that the field
 in terms of which the
partial fraction is formulated has to be a symmetry generator. But the complex analysis is blind to the
 subtle conformal nature of the field evaluated at $z_1$. It is clear that it can be {\it any field},
as long as (generically) the correlation function has {\it effectively} a {\it single channel}.  In
particular, it can be applied to the calculation of special correlators involving only primary
fields. Examples of such functions containing only Virasoro or parafermionic primary fields are
presented below.



\newsec{Correlators of Virasoro degenerate primary fields}

We will first consider the following correlation function of Virasoro degenerate primary fields:
$$\L\phi_{12}(z_1)\cdots \phi_{12}(z_n)\phi_{1,n+1}(z_{n+1})\R\eq$$
Given the fusion rule [\BPZ]
$$\phi_{12}\times \phi_{1r} = \phi_{1,r-1} + \phi_{1,r+1} \eq$$
we see that the insertion of the field $\phi_{1,n+1}$ in the last position effectively
selects a single channel, i.e., a single term contributes from each OPE. This is
most easily seen for the case $n=2$ corresponding to the three-point function: substituting 
$\phi_{12}\times \phi_{12} = \phi_{11} + \phi_{13}$ in the correlator and using the
orthogonality condition
$$\L\phi_{1r}(z_1)\phi_{1s}(z_2)\R= \delta_{r,s}\eq$$ it is clear  that it is  only the 
$\phi_{13}$ term does contribute. 

Recall that $\phi_{rs}$, with  $r,s$ integers, refers to a Virasoro primary field that contains a
singular descendant (hence the qualitative `degenerate') at level $rs$ and whose dimension $h_{rs}$
is linked to the central charge via a parameter
$t$ as follows:
$$\eqalign{ h_{rs}(t) &= \frac14(r^2-1)t+\frac14(s^2-1){1\over t}-\frac12(rs-1)\cr
c(t)&= 13-6\left(t+{1\over t}\right)\cr}\eq$$
It is convenient to set
$$a= {1\over 2t}\eq$$
In this notation, the dimension $h_{1n}$ takes the form
$$h_{1n}= \frac12(n^2-1)a -\frac12(n-1)\eq$$
so that the power of $(z-w)$ of the leading term of the family $\phi_{1m}$ in
the OPE of $\phi_{1r}\times \phi_{1s}$ is
$$h_{1r}+h_{1s}-h_{1m}= \frac12(r^2+s^2-m^2-1)a -\frac12(r+s-m-1)\eq$$

In order to lighten further the notation,  we will set
$$\phi_{1,r+1}\equiv\P_r\eq$$
and define the structure constants $c_{rs}$ as follows
$$\P_r(z)\P_s(w)\sim c_{rs} \P_{r+s} +\cdots\eq$$
Therefore, in terms of the minimal-model structure constants, the $c_{rs}$ are
$$c_{rs}= {C_{(1,r+1),(1,s+1)}}^{(1,r+s+1)}=  C_{(1,r+1),(1,s+1),(1,r+s+1)}\eq$$
the last expression being symmetric with respect to the interchange of any two pairs of the three
indices. 

Let us evaluate the four-point function
$$\L\phi_{12}(z_1)\phi_{12}(z_2)\phi_{12}(z_3)\phi_{14}(z_4)\R= \L
\P_1(z_1)\P_1(z_2)\P_1(z_3)\P_3(z_4)\R\eq$$
by the residue method.
The following function
$$F(z_1)= z_{12}^{-a}z_{13}^{-a}z_{14}^{5a-1}\L
\P_1(z_1)\P_1(z_2)\P_1(z_3)\P_3(z_4)\R\eq$$
turns out to an analytic function of $z_1$ (there are no poles).  Actually, it is simply a constant, as
the behavior as $z_1 \rw \y$ indicates, i.e.,
$$F(z_1)\sim  z_{1}^{-a}z_{1}^{-a}z_{1}^{5a-1}{1\over z_1^{3a-1}} \qquad (z_1\rw \y)\eq$$
where the last term is the contribution of the correlator {\it per se}, in which only $\P_1(z_1)$
contributes: $\P_1(z_1)\sim z_1^{-2h_{12}}$. This constant can be evaluated in many different ways and in
particular, in the limit $z_1\rw z_4$:
$$\eqalign{
\lim_{z_1\rw z_4}F(z_1) &= \lim_{z_1\rw z_4}z_{12}^{-a}z_{13}^{-a}z_{14}^{5a-1}
{c_{11}\over z_{14}^{5a-1}}\L
\P_2(z_2)\P_1(z_3)\P_3(z_4)\R\cr &=
\lim_{z_1\rw z_4}z_{12}^{-a}z_{13}^{-a}z_{14}^{5a-1}
{c_{11}\over z_{14}^{5a-1}}{c_{12}\over z_{23}^{-a}z_{24}^{4a-1}z_{34}^{4a-1}}\cr
&= {c_{11}c_{12}\over z_{23}^{-a}z_{24}^{5a-1}z_{34}^{5a-1}}\cr}\eq$$
Therefore, we have
$$ \L
\P_1(z_1)\P_1(z_2)\P_1(z_3)\P_3(z_4)\R=
{c_{11}c_{12}\over z_{12}^{-a}z_{13}^{-a}z_{23}^{-a}z_{14}^{5a-1} 
z_{24}^{5a-1}z_{34}^{5a-1}}
\eqlabel\fourpt$$

This computation can be easily generalized to the case where there is an arbitrary number of $\P_1$ fields
projected onto an appropriate $\P_n$ field enforcing the single-channel constraint:
$$\L \P_1(z_1)\cdots \P_1(z_n)\P_n(z_{n+1})\R= c_{11}c_{12}\cdots c_{1,n-1}
\prod_{1\leq i<j\leq n}{1\over z_{ij}^{-a}}\prod_{1\leq \ell\leq n}
{1\over z_{\ell, n+1}^{(n+2)a-1} }\eqlabel\genec$$
It is not difficult to verify that this correlator is solution of the $\phi_{12}$ singular-vector
differential equation.
Having computed our correlator without resorting to this differential equation, one could ask where does
the singular nature of $\phi_{12}$ enters, if it does at all. It is actually used right at the beginning,
in specifying the fusion rules. 

Still using the residue method, we can derive the following other generalization of the four-point
function  (\fourpt):
$$\eqalign{
\L
\P_1(z_1)\P_n(z_2) &\P_{n'}(z_3)\P_{n+n'+1}(z_4)\R\cr & = 
{c_{1,n+n'}c_{n,n'}\over z_{12}^{-na}z_{13}^{-n'a}z_{14}^{(3+n+n')a-1} 
z_{23}^{-nn'a}z_{24}^{(n^2+3n+nn')a-n}z_{34}^{(n'^2+3n'+nn')a-n'}}
\cr}\eq$$
A different ordering in the evaluation of the constant representing the 
intermediate meromorphic function forces the
relation
$$c_{1,n+n'}c_{n,n'}= c_{1,n}c_{n+1,n'}\eqlabel\etapedes$$
The  solution of this recursion relation reads 
$$c_{n,n'}= {c_{1,1}\cdots c_{1,n+n'-1}\over c_{1,1}\cdots c_{1,n-1}\, c_{1,1}\cdots
 c_{1,n'-1}}\eqlabel\etapedesol$$
which, as it should, is symmetric in both indices. 

We now want to stress that the factorized expression (\etapedes) was actually coded in the correlator
(\genec).
This correlator was computed  by
contracting all the
$\P_1$ fields from left to right. Equivalently, we could have stop this process at the $m$-th
one and contract the remaining $\P_1$'s from right to left up to the
$m'$-th. That yields (with $n=m+m'$):
$$\eqalign{ &[c_{11}c_{12}\cdots c_{1,m-1}] [c_{11}c_{12}\cdots c_{1,m'-1}]
\L\P_m\P_{m'}\P_{m+m'}\R\cr
&\quad \sim [c_{11}c_{12}\cdots c_{1,m-1}] [c_{11}c_{12}\cdots
c_{1,m'-1}] c_{m,m'}\cr}\eq$$ The comparison between the two
results yields directly (\etapedesol).

 Equation (\etapedesol) shows that all the constants $c_{n,n'}$ can be
calculated in terms of the $c_{1,n}$'s only.\foot{We stress 
that the $c_{n,n'}$'s form a particular
class of structure constants, first in that they pertain to 
the restricted $\{\phi_{1,r}\}$ algebra and
second, because they are those in front of the `highest field', 
i.e., the maximal value of $m$,  in the
OPE
$\P_r\times
\P_s\sim \sum_m\P_m$.} In order to calculate the structure constants $c_{1,n}$, we need to evaluate the
four-point function
$\L\P_1\P_n\P_1\P_n\R$; but in this case, the two channels in the OPE $\P_1\times \P_n$ do contribute,
which invalidates the applicability of the residue method.  This correlator can be evaluated by using the
singular-vector equations [\BPZ] or by using screening operators [\ref{Vl.S. Dotsenko and V.A. Fateev,
Nucl. Phys. {\bf B235} (1984) 312.}\refname\DFa, \ref{Vl.S. Dotsenko and V.A. Fateev, Nucl. Phys. {\bf
B251} (1985) 691.}\refname\DFb, \ref{Vl.S. Dotsenko and V.A. Fateev,  Phys. Lett. {\bf 154B} (1985)
291.}\refname\DFc]. But this will not be reconsidered here. Our main point  was to unravel the
factorization  (\etapedesol), as well as illustrating the residue method.

We can similarly write down rather directly the expression for all correlators of the form
$$\eqalign{
&\L \P_{r_1}(z_1)\P_{r_2}(z_2)\cdots \P_{r_n}(z_n)\P_{R}(z_{n +1})\R \cr
&= 
c_{s_1,r_2}c_{s_2,r_3}\cdots c_{s_{n-1},r_n}
\prod_{1\leq i<j\leq n}{1\over z_{ij}^{-r_ir_ja}}\prod_{1\leq \ell\leq n}
{1\over z_{\ell, n+1}^{r_\ell[(R+2)a-1]} }\cr}\eqlabel\genelo$$
with
$s_i=\sum_{j=1}^i r_j$ and $\sum_{i=1}^n r_i=R$.

Still more generally, we can also use the residue method to evaluate the correlators 
$$\L \phi_{21}(z_1)\cdots \phi_{21}(z_{r-1})\phi_{12}(z_r)\cdots
\phi_{12}(z_{r+s-1})\phi_{rs}(z_{r+s})\R \eq$$ since they  also involve a single channel. The
corresponding structures constants satisfy
$${C_{(r,s),(r',s')}}^{(r+r'-1,s+s'-1)}=  {C_{(r,1),(r',1)}}^{(r+r'-1,1)}\, 
{C_{(1,s),(1,s,)}}^{(1,s+s-1)}= c_{rr'}c_{ss'}\eq$$

Are these results completely surprising? From the Coulomb-gas representation point of view,
these expressions are somewhat trivial: these are the very correlators that do not require
the insertion of even a single screening operator (and for this reason they have not been
considered in [\DFa,\DFb]). A Coulomb-gas  correlation function without screening
is simply that of a collection of vertex operators, the result of which being quite simple
and well-known (see e.g., eq. (9.9) of [\CFT]). All the
$z_{ij}$ dependence of the correlators is recovered in this way, the remaining factors being
simply the structure constants. 

However we stress  that in our computation we do not require the free-field
representation. In this sense the present derivation is thus more fundamental, in spite of the fact that
it is applicable to a rather limited class of correlators. It also reveals a simple factorization
of some of the structure constants that may not have been obvious from other points of view, but whose
generality is by now quite transparent.

\newsec{Correlators in  parafermionic models}

\subsec{Reviewing  the $\ZZ_k$  parafermionic algebra}

The  $\ZZ_k$ parafermionic conformal algebra [\ZFa,\ref{A.B.
Zamolodchikov and V.A. Fateev, Sov. Phys. JETP  {\bf 63} (1986) 913.}\refname\ZFb] (see also 
[\ref{D. Gepner and Z. Qiu, Nucl. Phys.
{\bf 285} (1987) 423.}\refname\GQ,\ref{P. Jacob and P. Mathieu, Nucl. Phys. {\bf B 587} (2000)
514.}\refname\JM]
 is generated by 
 $k$
conserved holomorphic (and similar anti-holomorphic)
parafermionic fields
$\psi_n$, $n=0,1,\cdots , k-1$, with $\psi_0=I$ and $\psi^\dagger_n=
\psi_{k-n}$, with conformal dimension $h_{\psi_n}$ satisfying
$ h_{\psi_n} = h_{\psi_{k-n}}$. 
The set of 
conformal dimensions $\{h_{\psi_n} \}$ is an input of a parafermionic theory and for the
$\ZZ_k$ model, they read
$$  h_{\psi_n}={n(k-n) \over k}  \eqlabel\simdi$$
The $\psi_n$ are primary fields that form a closed algebra specified by the following
OPEs:
$$\eqalign{ 
\psi_n (z) \,\psi_{n'} (w) &\sim {{\bar c}_{n,n'}\over  (z-w)^{2nn'/k}}\;
 \psi_{n+n'} (w)  \qquad (n+n'<k) \cr
 \psi_n (z) \,\psi^\dagger_{n'} (w) &\sim { {\bar c}_{n,k-n'}\over 
(z-w)^{2{\rm min}(n,n')-2nn'/k} }\;
  \psi_{k+n-n'} (w)  \qquad (n+n'<k) \cr
\psi_n (z) \,\psi^\dagger_n (w) &\sim {1\over (z-w)^{2n(k-n)/k}}
\left[I+(z-w)^2 {2 h_{\psi_n} \over c}\, T(w) +\cdots\right]
\cr} \eqlabel\zkope$$
where the  central charge is fixed by associativity to be
$$c={2(k-1)\over (k+2)}  \eqlabel\zkcentral$$ 
The remaining OPE are obtained by conjugation and the condition
$${\bar c}_{n,n'}= {\bar c}_{k-n,k-n'}\eq$$
which implies, in particular, that 
${\bar c}_{n,k-n'}= {\bar c}_{k-n,n'}$.
The constants ${\bar c}_{n,n'}$  are assumed to be real. They are fixed by the associativity
conditions:
$$c^2_{n,n'}={\Gamma (n+n'+1) \Gamma (k-n+1) \Gamma (k-n'+1) \over 
\Gamma (n+1) \Gamma (n'+1) \Gamma (k-n-n'+1) \Gamma (k+1)}\quad\qquad (n+n'<k)
\eqlabel\ctestruc$$ Note that this yields ${\bar c}_{n,k-n}=1$.
They satisfy
$${\bar c}_{n,n'}= {\bar c}_{n,k-n-n'}= {\bar c}_{n',k-n-n'}\eqlabel\symdesc$$
which reflects the symmetry of the three-point function:
$$\L\psi_n(z_1)\psi_{n'}(z_2)\psi_{n+n'}^\dagger(z_3)\R= {C_{n,\,n',\,n+n'}\over 
z_{12}^{2nn'/k}  z_{23}^{2n'-2n'(n+n')/k}  z_{13}^{2n-2n(n+n')/k}}\eq$$





\subsec{$\ZZ_k$ parafermionic correlators and  their structure constants}

As a simple illustrative example of the application of the residue method to a parafermionic
correlator, let us first rederive the well-known expression for the three-point correlation
function
$$\L\psi_1(z_1)\psi_1(z_2)\psi_2^\dagger(z_3)\R\eq$$
Given the structure of the OPEs, we know that the function $$F(z_1)=
z_{12}^{2/k}z_{13}^{-4/k}\L\psi_1(z_1)\psi_1(z_2)\psi_2^\dagger(z_3)\R\eq$$ is a
meromorphic function of $z_1$ with a double pole at $z_3$ and no regular terms. 
Therefore, this function has to be of the form:
$$F(z_1) = {a_2\over z_{13}^{2}} + {a_1\over z_{13}}\eq$$
The first coefficient is 
$$a_2= \lim_{z_1\rw z_3} 
z_{13}^{2}z_{12}^{2/k}z_{13}^{-4/k}\L\psi_1(z_1)\psi_1(z_2)\psi_2^\dagger(z_3)\R\eq$$
Since we will be interested also in the second order term, we will need to consider also
the subleading contribution in the limiting value of the correlation function.
Recall that in the OPE of $\phi_a$ and $\phi_b$, the first two contributing terms in the
conformal family of
$\phi_c$ are
$$\phi_a(z)\phi_b(w)\sim {C_{abc}\over
(z-w)^{h_a+h_b-h_c}}\left[\phi_c(w)+{(h_a-h_b+h_c)\over 2h_c} (z-w)\,\d
\phi_c(w)+\cdots\right]\eq$$ In the present case, we need
$$\psi_1(z_1)\psi_2^\dagger(z_3) \sim {{\bar c}_{1,1}\over
z_{13}^{2-4/k}}\left[\psi_1^\dagger(z_3)+{z_{13}\over k-1}\d 
\psi_1^\dagger(z_3)+\cdots\right]\eq$$ Therefore
$$\lim_{z_1 \rw z_3} \L\psi_1(z_1)\psi_1(z_2)\psi_2^\dagger(z_3)\R=
\lim_{z_1 \rw z_3}{{\bar c}_{1,1}(-1)^{-2/k}\over z_{13}^{2-4/k} z_{23}^{2-2/k}}\left[
1+{2\over k} {z_{13}\over z_{23}}+\cdots \right]\eq$$ Only the leading term contributes
to the first coefficient, which  is thus 
$$a_2= {{\bar c}_{1,1}\over z_{23}^{2-4/k}}\eq$$
The other coefficient is given by
$$a_1= \lim_{z_1\rw z_3}{\d\over \d z_1}\left\{ 
z_{13}^{2}z_{12}^{2/k}z_{13}^{-4/k}\L\psi_1(z_1)\psi_1(z_2)\psi_2^\dagger(z_3)\R\right\}\eq$$
It is simple to check that it is equal to zero.  $F(z_1)$ has thus  a single
term;  the expression of the correlation function
under consideration is then
$$\L\psi_1(z_1)\psi_1(z_2)\psi_2^\dagger(z_3)\R= {{\bar c}_{1,1}\over z_{12}^{2/k}
z_{13}^{2-4/k}z_{23}^{2-4/k}}\eq$$
This is the correct  three-point function.

This computation  can easily be generalized  to the case where
there is an arbitrary number of $\psi_1$ factors :
$$\L\psi_1(z_1)\cdots \psi_1(z_n)\psi_n^\dagger(z_{n+1})\R = \prod_{1\leq
i<j\leq n}{{\bar c}_{1,i}\over z_{ij}^{2/k}} \prod_{1\leq
\ell\leq n} {1\over z_{\ell,n+1}^{2-2n/k} }\eqlabel\corNpts$$

Proceeding in a similar way, we can also compute the following correlator:
$$\eqalign{  \L\psi_1(z_1)\psi_n(z_2) &\psi_{n'}(z_3)\psi_{n+n'+1}^\dagger(z_{4})\R =
{{\bar c}_{n,n'}{\bar c}_{1,n+n'}\over
z_{12}^{2n/k} z_{13}^{2n'/k} z_{23}^{2nn'/k}}\cr
& \times {1\over   z_{14}^{2-2(n+n'+1)/k}
z_{24}^{2n(1-(n+n'+1)/k)}z_{34}^{2n'(1-(n+n'+1)/k)}}\cr}\eqlabel\cornnp$$
We now use this last expression to extract a first result on the structure
constants, by comparing the leading contribution of the limit $z_1\rw z_2$ of the
correlator:
$$\eqalign{\lim_{z_1\rw z_2} \L\psi_1(z_1)\psi_n(z_2)& \psi_{n'}(z_3)
\psi_{n+n'+1}^\dagger(z_{4})\R \simeq {{\bar c}_{1,n}\over z_{12}^{2n/k}}
\L\psi_{n+1}(z_2) \psi_{n'}(z_3)
\psi_{n+n'+1}^\dagger(z_{4})\R \cr
&\simeq {{\bar c}_{1,n}{\bar c}_{n+1,n'}\over
z_{12}^{2n/k}z_{23}^{2n'(n+1)/k}
z_{24}^{2(n+1)(1-(n+n'+1)/k)}z_{34}^{2n'(1-(n+n'+1)/k)}}\cr}\eq$$
with the same limit calculated directly from (\cornnp). This gives:
$${\bar c}_{1,n}{\bar c}_{n+1,n'}= {\bar c}_{n,n'}{\bar c}_{1,n+n'}\eq$$
precisely the same relation found for the minimal models and thus whose  solution is  again
exactly of the form (\etapedesol). The observation of this factorization 
seems to have first been made in
[\ref{P. Furlan, R.R. Paunov and I.V. Todorov, Fortschr. Phys. {\bf 40} (1992) 211.}\refname\FPT]. As already
hinted at, it is typical of monomial (i.e., single term) correlators.


In order to get the  complete expression for the coefficient ${\bar c}_{n,n'}$, we simply
need to evaluate
${\bar c}_{1,n}$.  This can also be done by the residue method.\foot{That was not the case for
 Virasoro primary-field correlators since not all $\P_1$ correlators do have a single
channel.   The difference here is that the
$\psi_n$'s  have a single OPE channel.}  For this we need to consider the
correlation function
$$G'=\L\psi_1(z_1)\psi_n(z_2) \psi_{1}^\dagger(z_3)\psi_{n}^\dagger(z_{4})\R\eq$$
with $n>1$.
As a function of $z_1$, 
$$F'(z_1)= z_{12}^{2n/k}z_{13}^{-2/k}z_{14}^{-2n/k}\L\psi_1(z_1)\psi_n(z_2)
\psi_{1}^\dagger(z_3)\psi_{n}^\dagger(z_{4})\R\eq$$ is a meromorphic function of $z_1$
with double poles at $z_3$ and $z_4$ and vanishing at infinity,  so that
$$F'(z_1) = {a_2\over z_{13}^{2}} + {a_1\over z_{13}}+{b_2\over z_{14}^{2}} +
{b_1\over z_{13}} \eq$$ 
A straightforward computation yields:
$$\eqalignD{
&a_2 = {z_{23}^{2n/k}\over z_{34}^{2n/k}z_{24}^{2n-2n^2/k}}\qquad\qquad 
&a_1 = -a_2\, {2n \over k}{z_{24}\over  z_{23}z_{34}}\cr
&b_2 = {c_{1,n-1}^2 \over z_{23}^{2-2n/k} z_{34}^{2n/k}z_{24}^{2(n-1)-2n^2/k}}\qquad\qquad 
&b_1 = b_2 {2\over (k-n+1)}{ z_{23}\over z_{24}z_{34}}\cr}\eq$$
from which we obtain:
$$\eqalign{ G'=   \left({z_{14} z_{23}  \over z_{12} z_{34} }\right)^{2n/k}&{1\over
z_{13}^{2-2/k} z_{24}^{2n-2n^2/k}} \left\{1-
{2n\over k}\,{ z_{13} z_{24}\over z_{23} z_{34} }\right.\cr&+\left. {\bar c}_{1,n-1}^2{ z_{13}^2
z_{24}^2\over z_{23}^2 z_{14}^2 }\left(1+ {2\over k-n+1}\, { z_{14} z_{23}\over
z_{24} z_{34} }\right)\right\}\cr}\eqlabel\quaptmoinsdebase$$
To extract the value of ${\bar c}_{1,n-1}$, we compare the expression of the correlator
evaluated directly in the limit $z_{12}\rw 0 $ and $z_{34}\rw 0$, which yields
$$\lim_{z_{1}\rw z_2\atop z_{3}\rw z_4}\L\psi_1(z_1)\psi_n(z_2)
\psi_{1}^\dagger(z_3)\psi_{n}^\dagger(z_{4})\R  \simeq
{{\bar c}_{1,n}^2\over  z_{12}^{2n/k}z_{34}^{2n/k}z_{24}^{2n-2n^2/k}}\eq$$
with $G'$ given in (\quaptmoinsdebase), evaluated in the same limit. One finds
that the leading terms do not match, $G'$ being more singular.  Hence, the
leading coefficient
 in $G'$ must cancel, which forces:
 $${\bar c}_{1,n-1}^2= {n(k-n+1)\over k}\eqlabel\ctepp$$ The substitution of this result into
(\etapedesol) leads to (\ctestruc).  Note, on the other hand, that the central charge is fixed by the
subleading term of the correlator $\L\psi_1(z_1)\psi_1(z_2)
\psi_{1}^\dagger(z_3)\psi_{1}^\dagger(z_{4})\R$.

\subsec{Some $\ZZ_k$ spin-field correlators and a relation between their structure constants}

\let\g\gamma

A similar analysis can be applied to spin-field correlation functions. Here we use the
notation $\s_i$ for the holomorphic part of $i$-th  spin field in the parafermionic theory,
with $\s_0=\s_k=I$. The fusion rules are 
$$\s_i\times \s_j= \s_{i+j}+\cdots\eqlabel\spinfu$$
where the dots stand for fields that are not parafermionic primary in the sense that their
charge
  is not linked to their dimension as for the spin field, for which they are
respectively 
$$ h_\ell = {\ell(k-\ell)\over 2k(k+2)}\qquad{\rm and}\qquad  q_\ell= \ell  \eq$$ 
These fusion rules are
obtained  from the coset realization $\suh(2)_k/\uh(1)$ [\GQ]. Let ${\tilde c}_{ij}$ be the structure
constant in front of the term which is given explicitly in (\spinfu), namely
$$\L \s_n(z_1)\s_{n'}(z_2)\s_{n+n'}^\dagger(z_3)\R = 
{{\tilde c}_{ij} \over z_{12}^{nn'b} z_{13}^{bn(k-n-n')} z_{23}^{bn'(k-n-n')} }\eq$$
where $$b={1\over k(k+2)}\eq$$

For the general  correlator of $n$ $\s_1(z_1)$ fields with $\s_n^\dagger$,
the residue method leads to the following simple expression:
$$\L\s_1(z_1)\cdots \s_1(z_n)\s_n^\dagger(z_{n+1})\R= {\tilde c}_{11}{\tilde c}_{12}\cdots {\tilde
c}_{1,n-1}
\prod_{1\leq i<j\leq n}{1\over z_{ij}^b}\prod_{1\leq i\leq n}{1\over z_{i,n+1}^{(k-n)b}}\eq$$

This again leads to the relation (\etapedesol) -- in tilde version --  for the structure constants.
 Obviously,  
this
relation could be derived as previously, by considering the correlator $\L
\s_1\s_n\s_{n'}\s_{n+n'+1}^\dagger\R$, which leads to (\etapedes) whose solution is the tilde version of
(\etapedesol). It is simple to check that the explicit expression found in  [\ZFa]  for the
${\tilde c}_{n,n'}$'s 
$${\tilde c}_{n,n'}^2= {\g(1)\g(n+n'+1)\g(k-n+1)\g(k-n'+1)\over \g(n+1)\g(n'+1)\g(k-n-n'+1)\g(k+1)}\eq$$
with 
$$\g(x)= \Gamma\left({x\over k+2}\right)\eq$$
indeed satisfies the factorization (\etapedesol).\foot{Note 
that the factorization by itself cannot lead
us to the above expression:  the exact value of ${\tilde c}_{1,n}$ is required  and it can be obtained
through  the calculation of the
correlator 
$\L \s_1\s_n \s_1^\dagger\s_n^\dagger\R$. But this cannot be evaluated by means of the residue
method because there are two contributing channels. In [\ZFa] it was evaluated  through the coset
representation. The coset approach is usually not a convenient way of computing correlation functions -
(cf. [\CFT], chap. 18); however, the simplicity of the present coset, namely $\suh(2)_k/\uh(1)$, 
 allows for a direct factorization of the correlators into a WZW piece and a free boson one.
The correlator could also be calculated from the parafermionic singular conditions. This calculation will
be reported elsewhere.}   This is at the roots of the curious similarity noted in [\ZFa]  between the
parafermionic structure constants and the spin-field ones.

\subsec{A $\ZZ_k^{(2)}$ correlator}

As already pointed out, in a parafermionic model, the conformal dimension of the parafermionic fields is
an input.  It is however constrained by the $\ZZ_k$ invariance itself, which requires $h_{\psi_n}=
h_{\psi_n^\dagger} $ and the monodromy invariance of the correlators. These constraints are satisfied
for
$$h_{\psi_n}^{(\beta)}= {\beta n(k-n)\over k}\eq$$ for any positive integer $\beta$.  However, the
underlying associativity conditions have to be checked anew for each value of $\beta$. The defining
OPEs are  given  by (\zkope) but with $(z-w)$ replaced by $(z-w)^\beta$ and
${\bar c}_{n,n'}\,\rw \,{\bar c}_{n,n'}^{(\beta)}$.  In the following, we denote the
corresponding parafermionic model as $\ZZ_k^{(\beta)}$, with
$\ZZ_k\equiv \ZZ_k^{(1)}$. 
The results of the associativity conditions for the case $\beta=2$ are been presented in [\ZFa] (cf.
their appendix A) and the special $\ZZ_3^{(2)}$ model is studied in [\ref{A.B. Zamolodchikov and V.A.
Fateev.  Theor. Math. Phys. {\bf 71} (1987) 163.}\refname\ZFc]. (We will report elsewhere on a detailed
analysis of the  $\ZZ_k^{(2)}$ models.) 

Let us consider a sample correlator of the $\ZZ_k^{(2)}$ model, namely $\L\psi_1\psi_1
\psi_{1}^\dagger\psi_{1}^\dagger\R$, from which we construct the meromorphic function
$${\tilde F}(z_1)= z_{12}^{4/k}z_{13}^{-4/k}z_{14}^{-4/k} \L\psi_1(z_1)\psi_1(z_2)
\psi_{1}^\dagger(z_3)\psi_{1}^\dagger(z_{4})\R\eq$$
It has poles or order 4 at $z_3$ and $z_4$ and behaves as $1/z_1^4$ as $z_1\rw\y$, meaning that there is no
analytic piece.  The computation of the two principal parts requires thus the knowledge of the first
three subleading terms in the OPE $\psi_1(z)
\psi_{1}^\dagger(w)$. That makes the computation much more involved than in the $\beta=1$ case. However,
there is a simple trick that allows us to avoid going so deeply inside the conformal block. Since
$F\sim 1/z_1^4$, we can  reduce the order of the two poles by two, at the price of adding a constant
term, by multiplying ${\tilde F}(z_1)$ by $z_{13}^{2}z_{14}^2$.  The transformed meromorphic function reads
thus
$${\tilde F'}(z_1)= z_{13}^{2}z_{14}^2 {\tilde F}(z_1)= {a_2\over z_{13}^2}+{a_1\over z_{13}}+{b_2\over
z_{14}^2}+{b_1\over z_{14}}+c_0\eq$$
For this computation we require the knowledge of only the first subleading term in the OPE 
$\psi_1(z)
\psi_{1}^\dagger(w)$. A  direct but rather long calculation gives the following expression for the
correlator:
$$\eqalign{ 
&{1\over (z_{12}z_{34})^{4/k} (z_{13}z_{14}
z_{23}z_{24})^{2-4/k} }\left\{ \left({\bar c}_{11}^{(2)}\right)^2 +
{z_{34}^2\over z_{23}z_{24}}-{4\over k}
 {z_{12}^2z_{34}^2\over z_{14}z_{13}z_{23}z_{24} } \right.\cr
&\left. \quad +  {z_{34}z_{24}\over z_{14}^2z_{23}^2 }(z_{12}z_{14}-z_{13}z_{24}) 
+  {z_{34}z_{23}\over z_{24}^2z_{13}^2 } (z_{14}z_{23}-z_{12}z_{13}) \right\}\cr}\eq$$
If we compare this expression in the limit $z_1\rw z_3$ with the correlator evaluated directly in
this same limit, we find that for $c$ expressed in terms of a parameter $\lambda$ as
$$c={4(k-1)\lambda(k+\lambda-1)\over (k+2\lambda)k+2\lambda-2)}\eq$$
the structure constant ${\bar c}_{11}^{(2)}$ becomes
$$\left({\bar c}_{11}^{(2)}\right)^2 = {2(k-1)(\lambda+1)(k+\lambda-2)\over k\lambda(k+\lambda-1)}\eq$$
in agreement with [\ZFc].  To check that the  $\ZZ_k^{(2)}$ central charge is unconstrained, we need
to compare the subleading terms.

Here again the constants ${\bar c}_{n,n'}^{(2)}$ have the factorization property (\etapedesol), so that
all constants can be expressed in terms of ${\bar c}_{1,n}^{(2)}$, whose determination requires the
evaluation of 
$\L\psi_1\psi_n
\psi_{1}^\dagger\psi_{n}^\dagger\R$.  Again the calculation can be reduced to the evaluation of poles
of order two. 
The expression of all the  structure constants ${\bar c}_{n,n'}^{(2)}$ are given in [\ZFc]. Curiously,
${\bar c}_{n,n'}^{(2)}$ has another factorization, namely as
$${\bar c}_{n,n'}^{(2)} = {\bar c}_{n,n'}d_{n,n'}(\lambda)\;, \quad
\qquad{\rm with}\qquad d_{n,n'}(1)= {\bar c}_{n,n'}\equiv {\bar c}_{n,n'}^{(1)}\eq$$
whose origin appears somewhat mysterious.

The structure constants for the models  $\ZZ_k^{(\beta>2)}$ can be computed in the same manner. 
Yet there exits no results concerning these theories. 

Coming back to methodological aspects, with  the computation presented in this subsection, we wanted
to emphasis that the evaluation of the principal part can be substantially simplified by an
appropriate modification of the integer powers of the prefactor multiplying the correlator.  Here,  poles of
order 4 have been transformed into poles of order 2.  This is not a purely technical issue. The determination of
higher subleading terms in the OPE $\psi\time \psi^\dagger$ requires the knowledge of the chiral algebra
underlying the $\ZZ_k^{(2)}$ models. More precisely, the required information can be extracted from various
associativity constraints  but it is clear that knowing at least the gross features of the underlying chiral
algebra is useful. For the $\ZZ_k^{(1)}$ model, it is the $WA_{k-1}$ algebra.  However, the chiral algebra of
the $\ZZ_k^{(2)}$ models is not known.


\newsec{Conclusion}

We have thus exposed, in its full generality, a  method for calculating special CFT correlators 
based an a direct application of the residue calculus. Let us first summarize the method and restate the limits
of its applicability.

We consider the correlator $\L A(z)\prod_i B_i(z_i)\R$ as a function of $z$. The first
step is to transform it into a meromorphic function of $z$. This is already guaranteed if the OPEs
$A(z)B_i(z_i)$ are Laurent series, which requires that the conformal dimensions of all the  fields
that appear in the singular terms of the OPE $A(z)B_i(z_i)$, as well as $A$ and $B_i$, differ by
integers.   This situation pertains to correlators of  chiral-algebra generators (in which case all the
fields have integer dimension). However, when $A$ is not a symmetry generator, the above condition is
rarely verified.  Generically, the OPE $A(z)B_i(z_i)$ contains a number of channels associated to primary
fields whose dimensions do not differ by integers. In such a case, one can still construct a
meromorphic function if either there is a single channel, or more generally, if a single channel, say
$C_i$,  contributes to the correlation function. This, however,  does not ensure that
$h_A+h_{B_i}-h_{C_i}$ is integer, i.e., that the OPE projected onto the $C_i$ channel is a Laurent
series. But the cure at this point is simple: one multiplies the OPE by a fractional
power of
$z-z_i$ suitably chosen to eliminate the algebraic singularity, leaving thus  pole-type
singular terms. In other words, when the effective single-channel requirement is satisfied, the correlator
can be transformed into a meromorphic function of $z$ by multiplying it by an appropriate factor
$\prod_i(z-z_i)^{d_i}$. This meromorphic function is then expanded in partial fractions, that is, as the
sum of the principal parts at the different $z_i$, including possibly the principal part at
infinity.

This method is certainly not new and we
try in appendix B to trace it back in the literature.   However, its explicit
spelling out as well as the identification of its inherent limitations appear to be new.  The method has
been illustrated here with various examples, including some 
 correlators that do not involve conserved currents.   As a practical application, we have
worked out a detailed derivation 
of the
parafermionic structure constants. 




\appendix{A}{A three-point function computed from the infinite series}

In this appendix, we consider the calculation of the three-point function $\L T(z_1)T(z_2)T(z_3)\R$ using
the brute force infinite-series method, where one OPE is replaced by its full infinite series
and show how this series can be summed exactly to reproduce the simple three-point function expression.

The long road  computation  will be compared to the few step computations that results from the
application of the conformal Ward identity:
$$\eqalign{
&\L T(z_1)T(z_2)T(z_3)\R  = \L\contract{T(z_1)T}(z_2)T(z_3)\R+ \L T(z_2)\contract{T(z_1)T}(z_3)\R\cr
&\quad = \left\L\left({c/2\over z_{12}^4}+{2T(z_2)\over z_{12}^2} + {\d T(z_2)\over z_{12}}\right)
T(z_3)\right\R + 
\left\L T(z_2)\left({c/2\over z_{13}^4}+{2T(z_3)\over z_{13}^2} + {\d T(z_3)\over z_{13}}\right)
\right\R\cr &\quad = {c\over z_{12}^2 z_{23}^4}-{2c\over z_{12} z_{23}^5}+{c\over z_{13}^2
z_{23}^4}+{2c\over z_{13} z_{23}^5} =  {c\over z_{12}^2 z_{13}^2z_{23}^2}\cr}\eqlabel\tttco$$

Let us now turn to the expression that results from the  replacement of the OPE $T(z_1)T(z_2)$ by its
infinite series:
$$\eqalign{T(z_1)T(z_2) & =  \contract{T(z_1)T}(z_2) + (T(z_1)T(z_2))\cr
&= \left({c/2\over z_{12}^4}+{2T(z_2)\over z_{12}^2} + {\d T(z_2)\over z_{12}}\right)+ \sum_{n=0}^\y
 {z_{21}^n\over n!} (TT^{(n)})(z_1)\cr}\eqlabel\ttopr$$
where $T^{(n)}= \d^n T$ and $(T(z_1)T(z_2))$ stands for the normal ordering of the product $T(z_1)T(z_2)$,
defined as
$$(AB)(z)= {1\over 2\pi i}\oint {dx\over x-z} A(x)B(z)\eq$$
The substitution of (\ttopr) into the three-point function yields 
$$
\L T(z_1)T(z_2)T(z_3)\R  = {c\over z_{12}^2 z_{23}^4}-{2c\over z_{12} z_{23}^5} +
 \sum_{n=0}^\y
 {z_{21}^n\over n!} \L
 (TT^{(n)})(z_1)T(z_3)\R\eqlabel\sada$$
In the last three-point function, we have to find the term proportional to 
$c\, z_{13}^{-6-n}$ which is the only contributing one; it is obtained by standard methods (see e.g.,
[\CFT]): in 
$$\contract{T(z_3)\,(T}T^{(n)})(z_1)=  {1\over 2\pi i}\oint {dx\over x-z_1}\left\{
\left(\contract{T(z_3)T}(x) \right)T^{(n)}(z_1)+  T(x)
\left(\contract{T(z_3)T^{(n)}}(z_1)\right) \right\}\eq$$ only the first piece contributes  to
$cz_{31}^{-6-n}$ and its different contributions add up $ (6+n)(3+n)!/12$. Therefore, we have
$$\L(T(z_1)T(z_2))T(z_3)\R= {c\over 12}\sum_{n=0}^\y {(3+n)!(6+n)\over n!}
{z_{12}^n\over z_{13}^{n+6}}\eq$$
The infinite series can be summed as follows:
$$\eqalign{\sum_{n=0}^\y &(3+n)(2+n)(1+n)(6+n)
{z_{12}^n\over z_{13}^n}  = {1\over z_{12}^5}\,\d_2\, z_{12}^6 \,\d_2^3\sum_{n=0}^\y
{z_{12}^{n+3}\over z_{13}^n}\cr
&= {1\over z_{12}^5}\,\d_2\, z_{12}^6\, \d_2^3
\left({z_{12}^3\over 1-z_{12}/z_{13}}\right) = {1\over z_{12}^5}\,\d_2\, z_{12}^6\, \d_2^3
\left({z_{12}^3z_{13}\over z_{23}}\right)\cr
&= {1\over z_{12}^5}\,\d_2\, z_{12}^6 \left({-6z_{13}^3\over z_{23}^4}\right)= {36z_{13}^3\over
z_{23}^4}+{24z_{13}^3z_{12}\over z_{23}^5}\cr}\eq$$
so that
$$\L(T(z_1)T(z_2))T(z_3)\R= {c\over z_{13}^2z_{23}^5}(3z_{23}+2z_{12}) =
{c\over z_{13}^2z_{23}^5}(z_{23}+2z_{13})\eq$$ The substitution of this expression into (\sada)
reproduces (\tttco).


\appendix{B}{Correlators as meromorphic functions through the literature}

As applied to correlators containing symmetry generators, the residue method could be traced back to
the pioneer work of Belavin, Polyakov and Zamolodchikov [\BPZ]. These authors explicitly consider the
correlator $\L T(z)\prod_i\phi_i(z_i)\R$ as a meromorphic function of $z$ with poles at $z_i$ whose
residues are fixed by the conformal properties of the field $\phi_i(z_i)$ (cf. the discussion after their 
eqs (2.8) and (3.3)). The method is also used in [\Zam]. In this  seminal paper, Zamoldchikov has
launched the exploration of extended conformal algebras through the study of the associativity conditions
of a number of cases containing a single extra symmetry generator. Correlation
functions are computed by considering only the singular terms in the OPEs. 
There is again an explicit reference to the
residue calculus (cf. eqs (2.1)-(2.4)) that suggests an underlying complex-analysis interpretation of the
exposed computations.  But note that such an approach is not mandatory since the calculation method  could
be justified by means of the  Ward identities associated to these extra conserved currents.

Meromorphicity is the central theme of Goddard's proposed formalization of conformal
field theory [\ref{P. Goddard, {\it Meromorphic conformal field theory}, in {\it Infinite dimensional Lie
algebras and groups}, ed. V. Kac, World Scientific (1989) 556.}\refname\God].  However, the meromorphic
point of view here is not implemented at the level of computing correlation functions.  In this context,
meromorphicity is used to establish locality which in turn becomes the corner-stone property for the
construction of the conformal field theory.  This formal procedure has been much developed in a sequel
work with Gaberdiel [\ref{M. Gaberdiel and P. Goddard, Commun.Math.Phys. {\bf 209} (2000)
549.}\refname\GGod].  Here the emphasis is placed  on the  reconstruction the whole theory out of
meromrophic amplitudes, via an approach inspired by the early works in dual models where the space of
states was originally built out of the conjectured dual amplitudes.  In this very paper however, the
residue method is explicitly invoked for the calculation of some correlators involving symmetry
generators.  In that regard, they present some results of Frenkel and Zhu [\ref{I.B. Frenkel and Y. Zhu, Duke
Math. J., {\bf 66} (1992) 123.}\refname\FZu], which have devised a nice combinatorial method for handling
such correlators (involving an arbitrary number of either $T$ factors  or affine Lie algebra generators
-- summarized in sections 5(b) and 5(c) of [\GGod]).  But again, the residue method is not systematized and its
applications are restricted to the elimination of conserved currents in correlation functions.  

It is in the context of a non-meromorphic theory that the partial fraction expansion technique has been 
mentioned in the most explicit way, namely as a natural tool for evaluating the parafermionic
correlator
$\L \psi_1\cdots  \psi_1 \psi_1^\dagger\cdots  \psi_1^\dagger\R$ [\ZFa] (cf. the discussion between eqs
(3.9)-(3.12)). In this paper, the authors also give the value of the parafermionic structure constants.
Even though there are no indication concerning the way these have been computed, it is natural to guess
that their calculations have been done roughly along the lines presented here. 


A variant of the partial fraction expansion method has been used explicitly in [\FPT] to work out 
in  detail the associativity conditions of the $\ZZ^{(2)}$ models (cf. their section 3B).  The approach used
there is superficially a little more complicated that the one presented here in that the four-point
correlation functions are transformed into meromorphic functions (actually, into polynomials) of the cross
ratio instead of functions of the position of one field. This procedure prevents an immediate generalization
to higher-point functions.

As applied to correlators that do not contain symmetry generators, we found afterwards a single
reference to the residue method: this is in appendix E of [\BPZ] (cf. eqs (E.14)-(E.17)).  It is used
there to calculate the Ising correlator $\L\psi(z)\sigma(z_1)\cdots \mu(z_{2M} ) \R$ where $\psi$ is the
Ising fermion and $\sigma$ and $\mu$ are respectively the spin and disorder fields. The correlator is
transformed into a meromrophic function of $z$ as follows:
$$F(z) = \left( \prod_{i=1}^{2M} (z-z_i)^{1/2}\right) \L\psi(z)\sigma(z_1)\cdots \mu(z_{2M}\R\eq$$
Actually, all singularities in $z$ are removed by this transformation since
$$\psi(z)\sigma(w)\sim{\mu(w)\over (z-w)^{1/2} }\, \qquad \quad
\psi(z)\mu(w)\sim{\sigma(w)\over (z-w)^{1/2}}\eq$$ Observe that these OPEs have a single channel. 
$F(z)$ is thus an 
 analytic function. The form of this analytic function is determined by its behavior at infinity.
Since
$F(z)\sim z^{M-1}$ as $z\rw \y$, it is necessarily a polynomial of order $M-1$, which can be written as
$$F(z)= \sum_{k=1}^{M-1}(z-z_{2M})^k g_k(z_i)\eq$$
(disregarding the dependence upon the anti-holomorphic variables).\foot{The choice of the expansion
variable, here $z-z_{2M}$, is conventional; another $z_j$ could have be chosen instead of $z_{2M}$ and the
power series could also have been written in powers of $z$ simply.  Writing the expansion in terms of  a
difference between two variables has the advantage of taking care of translation invariance. Note that there
is a misprint on the value of the upper limit of the sum in (E.16) of [\BPZ].} The coefficients $g_k$ are
determined by enforcing the correlator to be a solution of the singular-vector differential equation of the
free fermion ($=\phi_{21}$). Quite interestingly, the coefficient $g_0$ gives the value of the correlation
function without the fermion.

More recently, Dotsenko [\ref{Vl.S. Dostsenko, lectures at LPTHE (unpublished) and informal lecture at
the workshop {\it Quantum integrability}, CRM, May 2000; Vl.S.Dotsenko, J.L.Jacobsen and M.Picco,
{\it Parafermionic Theories}, to be published.}\refname\Dotse] has devised a nice way of
handling the parafermionic computations by a method which is close in spirit to the computation of [\BPZ]
just described,  hence to the partial fraction expansion. Applied  to four-point functions with three
points fixed at the special values 0, 1, $\y$, the idea is to factor out the branching or poles
singularities of the resulting $z$ function and then determine the remaining  polynomial in $z$ that
completes the correlator by considering successively the correlator in the limits where $z$ approaches the
three fixed points.  A detailed application of this method is presented in the following appendix.

\appendix{C}{The Dotsenko's method to test associativity}

We will illustrate the method initiated by Dotsenko by reconsidering the $\ZZ_k$ parafermionic
correlator
$\L\psi_1(z_1)\psi_n(z_2) \psi_{1}^\dagger(z_3)\psi_{n}^\dagger(z_{4})\R$. We fix three points
at the standard values 0, 1 and $\y$ and use the following convention for a `prime correlator':
$$\L X\,A(\y)\R'\equiv \lim_{z_n\rw \y} z_n^{2h_A}\L X\,A(z_n)\R \eq$$
In particular, we have
$$\L \psi_n(z) \psi_{n}^\dagger(\y)\R'= 1\; , \qquad \quad \L \psi_n(z)\psi_n(1)
\psi_{n+n'}^\dagger(\y)\R'= {{\bar c}_{n,n'}\over (z-1)^{2nn'/k}}\eq$$
 We consider thus
$ \L\psi_1(0)\psi_n(z) \psi_{1}^\dagger(1)\psi_{n}^\dagger(\y)\R$. Regarded as a function of $z$,
this has singularities at 0 and 1, that is
$${\cal G}(z)\equiv \L\psi_1(0)\psi_n(z) \psi_{1}^\dagger(1)\psi_{n}^\dagger(\y)\R' = {P_n(z)\over
z^{2n/k}(z-1)^{2-2n/k}}\eqlabel\dots$$
(we do not care about the phases which all cancel at the end). $P_n$ is a polynomial of order
$n$.  In other words, by multiplying the correlator by the prefactor $z^{2n/k}(z-1)^{2-2n/k}$, we
transform it into a meromorphic function, which turns out to be analytic.

The order of $P_n$ is fixed by considering the limit where $z\rw \y$, where we readily see
that
${\cal G}(\y)=1$, i.e.,
$$\lim_{z\rw \y}{\cal G} \sim \lim_{z\rw \y} \L \psi_1(0) \psi_{1}^\dagger(1)\R \L \psi_n(z)
\psi_{n}^\dagger(\y)\R' = 1\eq$$This implies that $n=2$ and, in addition, that the coefficient of the term
$z^2$ is 1, that is
$$P_2(z)= a_0+a_1 z+z^2\eq$$
Thus, in order to completely determine the correlator, we only have to fix these two constants. 

Consider first the limit where $z\rw 0$, keeping track of the first subleading term. The correlator
${\cal G}(z)$ becomes then
$$\eqalign{ \lim_{z\rw 0}{\cal G}(z) &= {{\bar c}_{1,n}\over z^{2n/k} } \L\psi_{n+1}(z)
\psi_{1}^\dagger(1)\psi_{n}^\dagger(\y)\R - {z\over n+1}{{\bar c}_{1,n}\over z^{2n/k} }
\L\d\psi_{n+1}(z)
\psi_{1}^\dagger(1)\psi_{n}^\dagger(\y)\R\cr &= {{\bar c}_{1,n}^2\over z^{2n/k} (z-1)^{2-2(n+1)/k}
}\left\{1+{2(k-n-1) z\over k(n+1)(z-1)}\right\}\cr &\simeq {{\bar c}_{1,n}^2\over z^{2n/k}} \left\{1+
{2k(k-n-1) z\over k(n+1)}\right\}\cr}\eq$$ This is to be compared with the expansion of the rhs of
(\dots): 
$$\lim_{z\rw 0} \left\{{a_0+a_1 z+z^2\over
z^{2n/k}(z-1)^{2-2n/k}}\right\}\simeq {1\over z^{2n/k}}\left[a_0+z\left(a_1+a_0 {2(k-n)\over
k}\right)\right]\eq$$ This yields
$$a_0= {\bar c}_{1,n}^2\, , \quad\qquad a_1=-{2a_0\over n+1}\eqlabel\aaa$$

Next we consider the limit $z\rw 1$. Here it will suffices to keep only the leading term. We have thus
$$ \lim_{z\rw 1}{\cal G}(z) \simeq {{\bar c}_{1,n-1}\over (z-1)^{2-2n/k} } \L\psi_{1}(0)
\psi_{k-1}^\dagger(1)\psi_{n}^\dagger(\y)\R \simeq {{\bar c}_{1,n-1}^2\over (z-1)^{2-2n/k} }$$
which is to be compared to 
$$\lim_{z\rw 1} \left\{{a_0+a_1 z+z^2\over
z^{2n/k}(z-1)^{2-2n/k}} \right\}\simeq {a_0+a_1+1\over
(z-1)^{2-2n/k}}\eq$$
That forces $$a_0+a_1+1 =  {\bar c}_{1,n-1}^2\eq$$ which together with (\aaa) yields
the recursion relation
$$ {\bar c}_{1,n-1}^2 =  {\bar c}_{1,n}^2\left({n-1\over n+1}\right) +1\eqlabel\recu$$
This relation can be solved without knowing the explicit value of ${\bar c}_{11}$ but simply by
enforcing the 
 condition
$${\bar c}_{1,n} = {\bar c}_{1,k-n-1}\eq$$
This readily implies 
$$ {\bar c}_{1,k-n}^2 =  {\bar c}_{1,k-n-1}^2\left({n-1\over n+1}\right) +1\eqlabel\recuu$$
On the other hand, replacing $n$ by $k-n$ in (\recu) yields 
$$ {\bar c}_{1,k-n-1}^2 =  {\bar c}_{1,k-n}^2\left({k-n+1\over k-n-1}\right) +1 \eqlabel\recuuu$$
Eliminating ${\bar c}_{1,k-n}^2$ from the last two equations lead to an algebraic relation for
${\bar c}_{1,k-n-1}^2$ which gives
$${\bar c}_{1,k-n-1}^2 = {\bar c}_{1,n}^2 = {(n+1)(k-n)\over k}\eq$$
and we recover the expression of ${\bar c}_{1,n}$ obtained previously in (\ctepp).

At first sight, it seems that the applicability of this method depends critically upon the
fact that the field evaluated at $z$ is the conjugate of the one at infinity.  In that case, the
other two fields are conjugate of each other, which ensures that the fractional powers of $z$ cancel in
the denominator as $z\rw\y$.  That certainly ensures the polynomial character of
$P_n$ as in the large $z$ behavior $\psi_n(z)$ is simply projected onto $\psi_n^\dagger(\y)$ and
$\L \psi_n(z)  \psi_n^\dagger(\y)\R'=1$. But consider instead the correlator
$$\L\psi_1(0)\psi_n(z) \psi_{n}^\dagger(1)\psi_{1}^\dagger(\y)\R'= {Q(z)\over 
z^{2n/k}(z-1)^{2n-2n^2/k}}\eqlabel\surt$$
It is not clear at once that $Q(z)$ is polynomial here. The point however is that  the large $z$
limit of
$\L A(0) B(z)C(1)D(\y)\R'$ is actually given by $z^{h_B+h_D-h_E}/z^{2h_B}$ where $E$ is the single
contributing field appearing in the OPE of
$B$ and $D$. Indeed, by considering at first the limit $z_4\rw\y$, we wash out the contribution
$z_{24}^{h_B+h_D-h_E}$ that needs to be reinserted at this point in order to get the right large $z$
behavior. The term $z^{-2h_B}$ simply corresponds to the large $z$ behavior of the $B$ field.  Returning
to our problem,  we need to compare the large
$z$ limit of the rhs of (\surt) with $z^{2-2n+2n^2/k-2n/k}\sim z^2$ which shows that $Q(z)$ is indeed a
polynomial of degree 2.

It should be clear from this example that the Dotsenko's method has the same intrinsic
limitations as the partial fraction expansion described in the main part of the article.  In particular,
it generically applies to  correlators that have a single contributing channel.

\vskip0.3cm
\centerline{\bf Acknowledgment}
We thank V.I. Dotsenko for very stimulating discussions that have triggered our interest for the subject
of this work. We would also like to thank  M. Gaberdiel (who pointed out to us [\GGod] and [\FZu]), Y.
Saint-Aubin  and G. Watts for useful comments and J.-F. Carrier and D. Huard for exploratory calculations
related to this work. We are also grateful to P. Furlan for pointing out [\FPT]. We acknowledge the financial
support of NSERC and FCAR.


\vskip0.3cm

\centerline{\bf REFERENCES}
\immediate\closeout\refs \vskip 0.5cm
  \message{References}\input references
\vfill\eject

\end